\documentstyle[emulateapj]{article}
\received{}
\revised{}
\accepted{}
\journalid{}{}
\articleid{}{}
\paperid{}
\cpright{}{}
\ccc{}
\lefthead{Matthews \& Wood}
\righthead{High-Latitude \HI\ in UGC~7321}

\begin{document}

\newcommand{\msun}{\mbox{${\cal M}_\odot$}}
\newcommand{\lsun}{\mbox{${\cal L}_\odot$}}
\newcommand{\kms}{\mbox{km s$^{-1}$}}
\newcommand{\HI}{\mbox{H\,{\sc i}}}
\newcommand{\mhi}{\mbox{${\cal M}_{HI}$}}
\def\etal{{\it et al.}}
\newcommand{\HII}{\mbox{H\,{\sc ii}}}
\newcommand{\NII}{\mbox{N\,{\sc ii}}}
\newcommand{\MgII}{\mbox{Mg\,{\sc ii}}}
\newcommand{\CIV}{\mbox{C\,{\sc iv}}}
\newcommand{\am}[2]{$#1'\,\hspace{-1.7mm}.\hspace{.0mm}#2$}
\newcommand{\as}[2]{$#1''\,\hspace{-1.7mm}.\hspace{.0mm}#2$}
\def\lsim{~\rlap{$<$}{\lower 1.0ex\hbox{$\sim$}}}
\def\gsim{~\rlap{$>$}{\lower 1.0ex\hbox{$\sim$}}}
\newcommand{\dark}{$M_{HI}/L_{B}$}

\title{High-Latitude \HI\ in the Low Surface Brightness Galaxy UGC~7321}

\vskip0.5cm
\author{L. D. Matthews\altaffilmark{1}}
\author{Kenneth Wood\altaffilmark{2}}

\altaffiltext{1}{Harvard-Smithsonian Center for Astrophysics, 60
Garden Street, MS-42, Cambridge, MA 02138 USA,
Electronic mail: lmatthew@cfa.harvard.edu}
\altaffiltext{2}{St. Andrews University, Dept. of Physics \&
Astronomy, North Haugh, St. Andrews KY16 9SS Scotland}

\singlespace
\tighten
\begin{abstract}
From the analysis of sensitive \HI\ 21-cm line observations, 
we find evidence for vertically extended \HI\ emission
($|z|\lsim$2.4~kpc) in the edge-on, low
surface brightness spiral galaxy UGC~7321.  
Three-dimensional modelling suggests that the \HI\ disk
of UGC~7321 is both warped and flared, but that 
neither effect can fully reproduce the
spatial distribution and kinematics of 
the highest $z$-height
gas. We are able to model 
the high-latitude emission as an
additional \HI\ component in the form of a ``thick
disk'' or ``halo'' with a FWHM$\sim$3.3~kpc. We find tentative evidence
that  the vertically extended gas declines in rotational
velocity as a function of $z$, although we are unable to
completely rule out models with constant $V(z)$.
In spite of the low star
formation rate of UGC~7321,
energy from supernovae may be sufficient to sustain this 
high-latitude gas. 
However,  alternative origins for this material, such as slow,
sustained infall, cannot yet be excluded.

\end{abstract}


\section{Background}
UGC~7321 is a nearby,  isolated Sd spiral galaxy seen  
nearly edge-on to our line of sight ($i\approx88^{\circ}$). The
distance to UGC~7321 is a matter of some uncertainty, but based on the
brightest resolved stars it is estimated to be $\approx$10~Mpc (see Uson
\& Matthews 2003),  and we adopt this value in the present
work. UGC~7321 exhibits a highly flattened, diffuse stellar disk with
no bulge component (Figure~1), and a very small intrinsic stellar 
scale height 
($h_{z}\sim$150~pc; Matthews 2000).
Its photometric properties have been studied by Matthews,
Gallagher, \& van Driel (1999), who found that it has
an intrinsically 
low optical surface brightness disk 
[$\mu_{B,i}(0)\approx$23.5 mag arcsec$^{-2}$], indicative of a low
current star formation rate. This is corroborated by its
low far-infrared luminosity ($L_{FIR}\approx7\times10^{7}L_{\odot}$), 
the 
nondetection of 21-cm radio continuum emission by Uson \& Matthews 
(2003; $F_{cont}\le0.41\pm0.25$~mJy),  and the rather
modest H$\alpha$ emission in this galaxy
($L_{H\alpha}\sim1\times10^{40}$~erg~s$^{-1}$; see
Section~\ref{SN}). 
The high neutral gas content of
UGC~7321 relative to its blue luminosity 
(\dark$\sim$1.0 \msun/\lsun), together with its
low dust content (Matthews \& Wood 2001), moreover suggest that this
galaxy has been an inefficient star-former over most of its lifetime.

Uson \& Matthews (2003) 
recently obtained new aperture synthesis
imaging observations of UGC~7321 in the \HI\ 21-cm line 
using the Very Large Array (VLA)\footnote{The Very
Large Array of the National Radio Astronomy Observatory is a facility
of the National Science Foundation, operated under cooperative agreement
by Associated Universities, Inc.}. Details regarding these
observations, as well as the global and kinematic properties of the
neutral hydrogen in UGC~7321 are
described by those authors. With a total of
12.3 hours of on-source integration time, the VLA data for UGC~7321 are
among the most sensitive \HI\ observations of
edge-on spiral galaxies obtained to date. Since the observations also have
reasonably
good spatial and spectral resolution ($\sim16''$ and $\sim$5.2~\kms,
respectively), they are well-suited for structural and kinematic
studies of the \HI\ gas in this nearby system. Here we
use these data to explore the vertical distribution of the 
neutral hydrogen gas in UGC~7321. 
With the aid of three-dimensional models,
we report evidence that the structure and
kinematics of the \HI\ component in 
UGC~7321  are more complex than a simple, thin-disk
distribution. UGC~7321 has a cool
\HI\ disk that appears to be both warped and flared, but our
analysis 
also suggests the presence of additional vertically-extended emission
(Section~\ref{3Dmod}). In Section~\ref{discussion}
we briefly discuss possible origins for this high-latitude material.

\section{HI Halos of Galaxies}
\HI\ surveys of the Milky Way have revealed that the
distribution of its neutral hydrogen component is far more complex
than simply a cool, thin disk of material lying along the Galactic
midplane. The Galactic neutral medium instead contains both cool and warm
layers of differing scale heights 
(Lockman 1984), as well as a halo component extending to
substantial $z$-heights ($h_{z}\sim$4.4~kpc; Kalberla et al. 1998).

From sensitive aperture synthesis observations, 
high-latitude \HI\ gas has now also been detected in 
other nearby galaxies. For example,
using deep \HI\ observations of the nearby, edge-on Sbc spiral 
NGC~891, Swaters, Sancisi, \& van der Hulst (1997) reported evidence
for \HI\ gas extending to heights of at least 5~kpc above the plane. 
Based on three-dimensional modelling, these
authors argued that this emission is part of an \HI\ halo that lags in
rotation speed by 25-100~\kms\ relative to material in the
disk.\footnote{Benjamin 2002 points out 
that the high end of this range
appears to be inconsistent with H$\alpha$ measurements of this
galaxy.}  Similar
analyses for the moderate inclination spiral NGC~2403 led to the
suggestion that this galaxy also has a rotationally lagging \HI\ halo 
and anomalous \HI\ material that may be related
to Galactic high-velocity clouds (Schaap, Sancisi, \& Swaters 2000;
Fraternali et al. 2001,2002). In addition, high-latitude \HI\ 
features now have been seen in NGC~5775
and NGC~2613 (Irwin 1994; Irwin \& Chaves
2003). Interestingly, NGC~5775 shows evidence for an \HI\
halo, while NGC~2613 seems to show only multiple discrete
high-latitude features (Irwin \& Chaves 2003).

For NGC~891 and NGC~2403, Swaters et al. (1997) and
Schaap et al. (2000), respectively, postulated that
the \HI\ halo gas in these galaxies
is likely to have originated from a galactic fountain-type  mechanism
of the type described by Shapiro \& Field (1976) and Bregman (1980). 
This explanation seems  
plausible for these systems, since both are
bright galaxies with moderately high
star formation rates and significant H$\alpha$ and radio continuum emission. 
Still, the sample of nearby galaxies for which gaseous halos,
particularly \HI\ halos, have been detected remains quite small, 
and a number
of puzzles regarding their structure and origin remain. 

One
difficulty with the galactic fountain scenario for the origin of
high-latitude gas is that the energetics
required to raise material 
from the disk to $z$-heights of several kpc often appears to be much
larger than can seemingly be supplied by ordinary supernova events 
alone (e.g., Rand \& van der Hulst 1993;
Lee \& Irwin 1997). Moreover,
evidence is accumulating that the observed vertical velocity 
gradients in
halos are not entirely consistent with fountain predictions 
(Collins, Benjamin, \& Rand 2002).

In some cases, it has been suggested that 
extraplanar gas, including \HI, may 
have been acquired during interactions (e.g., Sancisi 1988) or
buoyed up by tidally-triggered instabilities.
Nonetheless, interactions seem an unlikely source for the halo 
gas in either NGC~891 or
NGC~2403, since both of these galaxies appear to be isolated.
One possible alternative for the origin of at least some fraction
of the high-$z$ \HI\ material
in galaxies like these
is that this gas is a remnant from an earlier evolutionary phase or
is accreted via a slow ``drizzle'' (e.g., Oort
1970). 
The interest in these types of scenarios is further fueled by the 
recent results from 
QSO absorption line experiments toward moderate redshift galaxies
(redshifts $\lsim$1) that
have shown these systems frequently are
surrounded by very extended gaseous halos with covering factors of near
unity (e.g., Lanzetta et
al. 1995; Chen et al. 2001b). The 
large spatial extents of these halos ($\sim$180 $h^{-1}$ kpc), 
their
presence around galaxies spanning all morphological types, and the inferred
correlations between halo size and galaxy mass (Chen et al. 2001b),
all hint that their 
origin may be tied to processes intrinsic to 
the galaxy formation process rather than
the details of current star
formation (see also Section~\ref{discussion} 
below). It follows 
that at least some remnant of  such halos still may
be present around many disk galaxies in the local universe,
particularly
if the halos are continually replenished.

One important way to
test whether at least some fraction of the high-latitude \HI\ emission
observed in spiral disks could be linked to the galaxy formation
process rather than a by-product of vigorous star formation, tidal
interactions,  or mergers
is to search for evidence of high $z$-height \HI\ emission
in isolated disk galaxies  with very low levels of current 
star formation. \HI-rich, low surface brightness (LSB) ``superthin''
spirals are just such
objects (e.g., Uson \& Matthews 2003). 
Since high $z$-height gas can be most
unambiguously detected in highly-inclined galaxies, we have chosen the
edge-on LSB spiral UGC~7321 for such an 
investigation.

\section{Results}
\subsection{The HI Total Intensity Map}
A total \HI\ intensity map of UGC~7321 is shown in
Figure~\ref{fig:mom0}, overlayed on an optical $R$-band image from
Matthews et al. (1999). It is apparent from this figure that 
even after accounting
for the finite resolution of the VLA beam,
the \HI\ emission in UGC~7321 is not exclusively 
confined to a very thin layer. If
the \HI\ disk had an intrinsic FWHM thickness less than or 
comparable to that of the stars (i.e., \lsim\as{5}{3}; Matthews 2000), 
the apparent thickness
as observed with a 16$''$ FWHM Gaussian 
beam should be $\sim$17$''$. Instead we find based on 
intensity slices parallel to the minor axis that
the \HI\ layer has a mean apparent thickness of $\sim24''$ FWHM. 
In addition, a number of filamentary features 
protrude to even higher $z$. That these features are not simply noise
is implied by the fact that the disturbances are evident in several
successive underlying contours (see also Figure~7 of Uson \& Matthews 2003). 

One-dimensional \HI\ intensity slices extracted perpendicular to the disk of
UGC~7321 at various projected radii consistently
show cores that are Gaussian in shape,
but with additional extended wings (Figure~\ref{fig:1Dcuts}).
These wings stretch to
$z$-distances of up to $\sim\pm50''$ ($\sim$2.4~kpc).
Qualitatively the profiles shown in  Figure~\ref{fig:1Dcuts}
resemble plots of the distribution of \HI\ volume density as a
function of $z$ in the Milky Way disk
(see Lockman 1984, his Figure~5). In the
case of the Milky Way, Lockman showed that the observed 
density profile could be modelled as the sum of two
Gaussian components, together
with a more diffuse and extended exponential component. After correcting for
underestimated low-$z$ emission and a distance to the Galactic Center of
$R_{0}$=8.5~kpc, Dickey \& Lockman (1990) give FWHM for the
Gaussian components of 212~pc and 530~pc, and a scale height for the
exponential component of 403~pc.
The nature of the vertical intensity cuts in UGC~7321 therefore
hints at the possibility of an 
analogous multi-component \HI\ disk structure in
this ``superthin'' LSB galaxy. 

While suggestive of an extended gas component, 
the vertical intensity cuts and total intensity map that we have derived for 
UGC~7321 give us only a projected density distribution and do not
permit us to assess at what galactocentric radius the vertically
extended material is located.
For example, it is feasible that 
warped or flared gas from the outer disk regions
seen along the line of sight is affecting the shape of the
wings of the observed
intensity profiles, and thus could 
mimic the signatures of a  multi-component disk.
More firmly establishing the
presence of a multiple-component \HI\ disk in UGC~7321 therefore requires
exploiting the full three-dimensional nature of the \HI\ data cube,
and 
utilizing velocity as well as spatial information. 
We undertake this exercise in the following subsections.

\subsection{Minor-Axis Position-Velocity Profiles}
One powerful way of probing the structure of vertically extended gas in
disk galaxies is to examine 
position-velocity (P-V) profiles  perpendicular to the disk.
In Figure~\ref{fig:modZPV} (right column) we plot such diagrams for UGC~7321
at several disk locations. In these panels,
\HI\ emission can be seen extending up to $\sim 50''$
($\sim$2.4~kpc) from the plane. At these and other disk locations, most,
although not all, of this vertically extended material has 
velocities within $\le$50~\kms\ of the systemic velocity.

In a galaxy disk viewed edge-on,
gas that is extended in $z$ and that is primarily 
clustered near the systemic velocity
can be a signature of flaring of the \HI\ layer (Sancisi \& Allen 1979). 
However, certain
characteristics of the high-latitude material in 
our present data argue against this explanation for
UGC~7321. One is the existence of 
asymmetries in these features on the 
$+z$ versus $-z$ sides of the disk (see
Figure~\ref{fig:1Dcuts} \& \ref{fig:modZPV}), as well on 
the east and west sides
(compare 
the panels at $r=\pm20''$ and $\pm40''$ in 
Figure~\ref{fig:1Dcuts}). For such a highly inclined galaxy, a 
flare should be expected to be largely symmetric about the
midplane. East-west asymmetries do not strictly exclude a flare, but
would imply that the midplane density is different on the
two sides of the disk. 
A second argument against a pure flare model 
is that the degree of disk flaring
required to explain the high-$z$ emission would produce
unique signatures in both the \HI\ total intensity map and 
in the individual
channel maps. As we show below, these features are not seen. 
Thus as we demonstrate in the next sections, 
while UGC~7321 does
exhibit manifestations of both warping and flaring in its \HI\ disk, 
neither of these effects alone appears to be 
sufficient to consistently explain the
morphology and kinematics of its highest-latitude \HI\ emission. 

\subsection{Three-Dimensional Galaxy Modelling\protect\label{3Dmod}}
In order to further probe the nature of the high-$z$-height
\HI\ emission in UGC~7321, 
we have computed full
three-dimensional (3-D) models of the galaxy and
compared these with the observed \HI\ channel maps, vertical P-V
profiles, and total \HI\ intensity map. 
For this analysis we have modelled the galaxy velocity field using 
the rotation curve for UGC~7321 derived by Uson \& Matthews
(2003), with the
approaching and receding sides of the disk averaged. 
Assuming a constant Gaussian velocity
dispersion for the gas of 7~\kms, we then computed the
\HI\ intensity and velocity distribution using 
a modified version of the radiative transfer and 
rotation curve modelling code
written by one of us (KW; see Matthews \& Wood 2001). These model cubes
could be subsequently 
analyzed in a manner identical to the real \HI\ data set.

For our present
analysis we have assumed that the \HI\ is optically thin and that its
density profile is Gaussian
in both the $r$ and $z$ directions. The assumption that the gas is
largely optically thin appears to be justified, since Uson \& Matthews
(2003) found evidence of only very mild self-absorption toward the
center of UGC~7321 (where gas densities should be highest, and
velocity crowding most extreme). Moreover, any optically thick
clumps are likely to be confined primarily to the 
midplane, whereas in the present paper we are mainly concerned with
probing the spatial and velocity structure of the higher latitude gas.

The choice of a Gaussian for the vertical intensity distribution is
motivated by  its
success in describing the bulk of the 
global, vertical \HI\ distribution of
other galaxies [e.g., the Milky Way; Lockman (1984)] and 
its success in fitting our present data. However, a sech$^{2}z$
distribution (as expected for an isothermal disk) has a very similar
shape to a Gaussian and would also be consistent with 
the data. This choice does not affect our conclusions.

For UGC~7321 we also find a Gaussian to provide a  
good representation of the radial \HI\ intensity 
distribution, as is commonly found in
late-type spirals (e.g., Krumm \& Salpeter 1980). Other forms for the radial
distribution (such as an exponential) provided significantly
less satisfactory fits.
In all of our models,  the \HI\ intensity in the $r$ 
direction has a scale length of 5.6~kpc (13.2~kpc FWHM) and
is truncated at a radius of $r_{max}$=11.8~kpc. All of our
model data cubes also have been convolved 
with a Gaussian in the spatial domain 
in order to match the 
resolution of the VLA data. We adopt an inclination
for UGC~7321 of $i=88^{\circ}$ and a position angle of 82$^{\circ}$.
All of our model parameters are quoted as their values
prior to convolution with the telescope beam.
For each of the classes of models we describe below, we
have explored a wide range of relevant parameter space, but limit
our current discussion to only the best-fitting model of each category. 
 
\subsubsection{Single-Component Disk Models}
\begin{figure}[t]
\plotfiddle{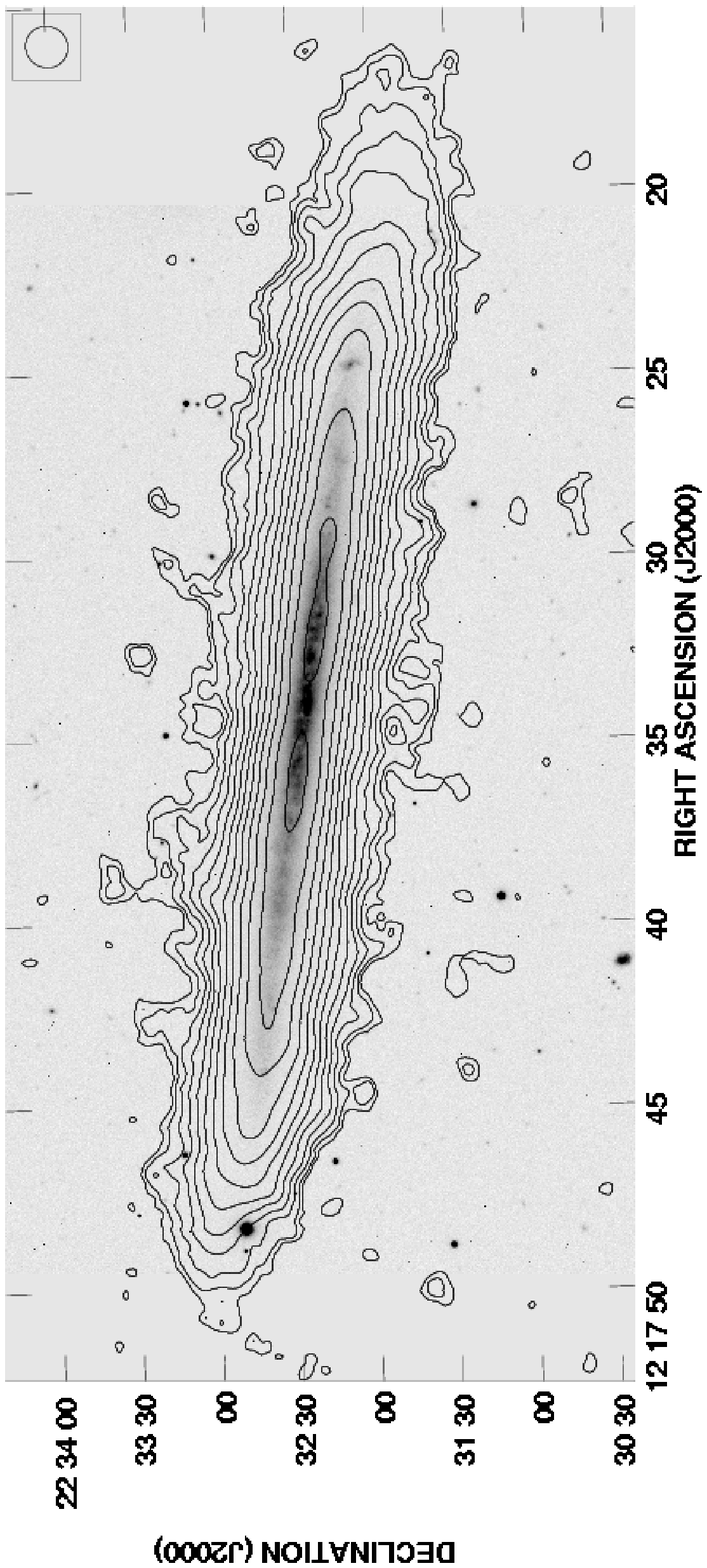}{6.0in}{-90}{70}{70}{-270}{420}
\figcaption{\HI\ total intensity map for UGC~7321 overlayed on an
$R$-band image from Matthews et al. 1999. Contour levels are
2,2.8,4,...44,64,88$\times(7.7\times10^{19}$) atoms
cm$^{-2}$. The FWHM beam size is $\sim$16$''$. 
\protect\label{fig:mom0}}
\end{figure}

\begin{figure}[t]
\plotfiddle{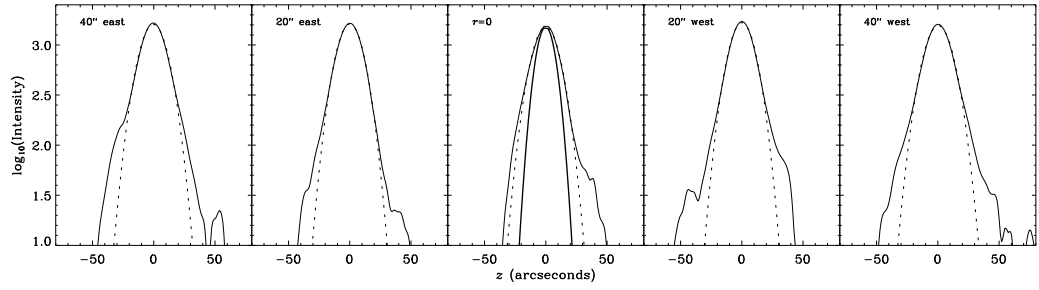}{8.0in}{90}{90}{90}{410}{-200}
\figcaption{\HI\ intensity profiles extracted perpendicular to the
midplane of UGC~7321 at several locations along the disk using the
data shown in Figure~\ref{fig:mom0} (thin solid lines). 
Single Gaussian
fits to the data are overplotted as dotted lines. FWHM of the fitted
Gaussians 
range from \as{22}{4}-\as{24}{4}. For comparison, the thick solid line
on the center panel shows a Gaussian with FWHM equal to the
synthesized beamwidth
($16''$). At all locations,
extended wings are visible in the data in excess of the
Gaussian models.
\protect\label{fig:1Dcuts}}
\end{figure}

The first class of model that we considered for UGC~7321 was that of a 
smooth, azimuthally symmetric Gaussian intensity distribution in both
the $r$ and $z$ directions (hereafter,
the ``smooth model''). 
We already know this model to be somewhat of an oversimplification,
since UGC~7321's gas disk is clearly is mildly warped
(Figure~\ref{fig:mom0}; see also Uson \& Matthews 2003). 
However, it serves as an important
baseline for comparison with more complex models.

\begin{figure}
\plotfiddle{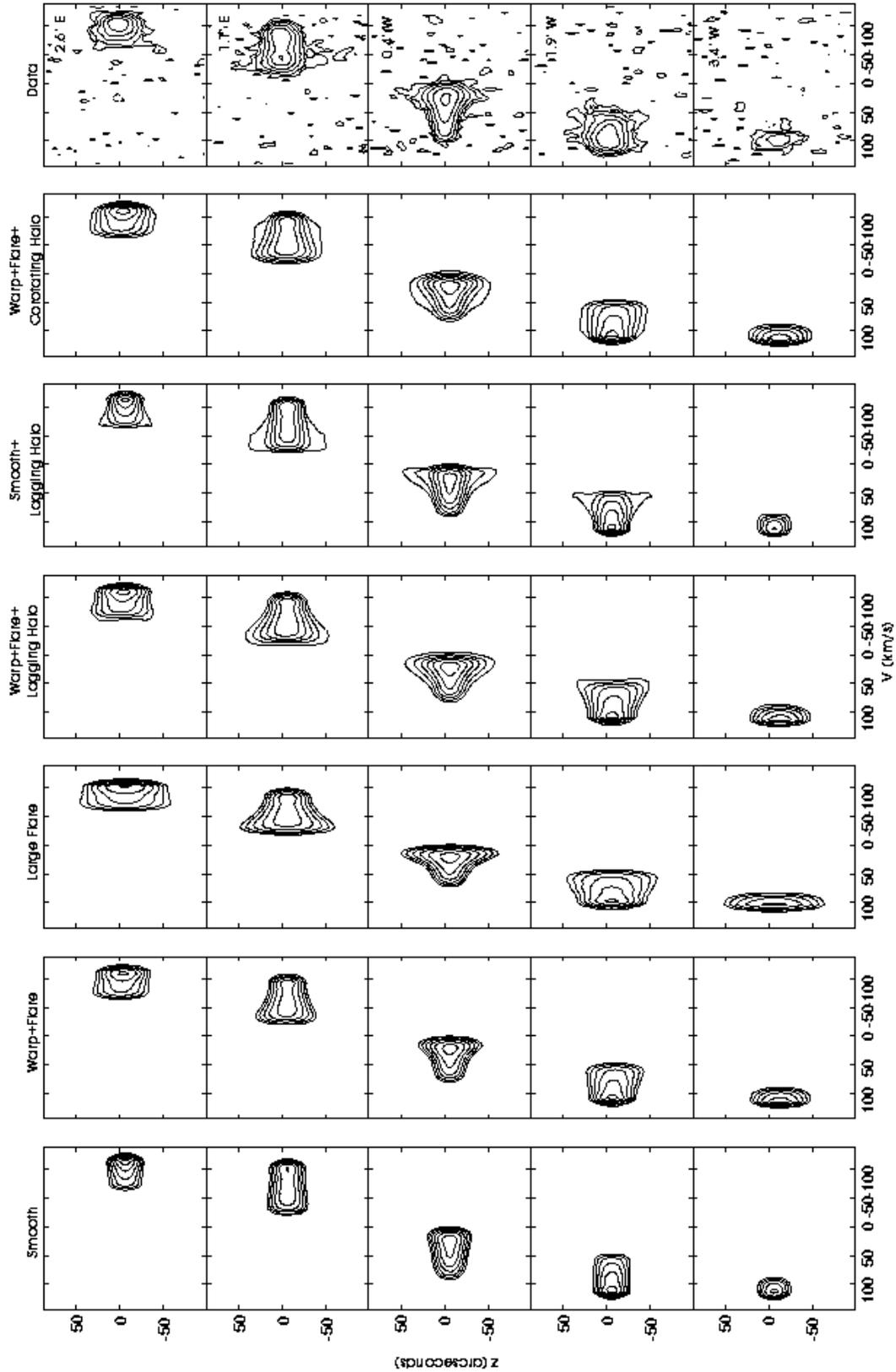}{8.0in}{0}{90}{90}{-300}{-60}
\figcaption{Position-velocity profiles extracted parallel to the minor
axis of UGC~7321 at several disk locations (right column) and at the
corresponding locations in various 
galaxy models discussed in Section~\ref{3Dmod}. Axes are velocity
(in \kms\ with respect to the systemic velocity) and the distance from the
disk midplane (in arcseconds). Contour levels are
$-0.2$,0.2,0.4,0.8,....12.8$\times$(0.39 mJy
beam$^{-1}$). The lowest contour
level is $\sim2\sigma$.
The profiles shown in each row were
extracted at \am{1}{5} intervals: 
\am{2}{6} and \am{1}{1} east of the minor axis; \am{0}{4}, \am{1}{9},
and \am{3}{4}
west of the minor axis.
\protect\label{fig:modZPV}}
\end{figure}

\begin{figure}
\plotfiddle{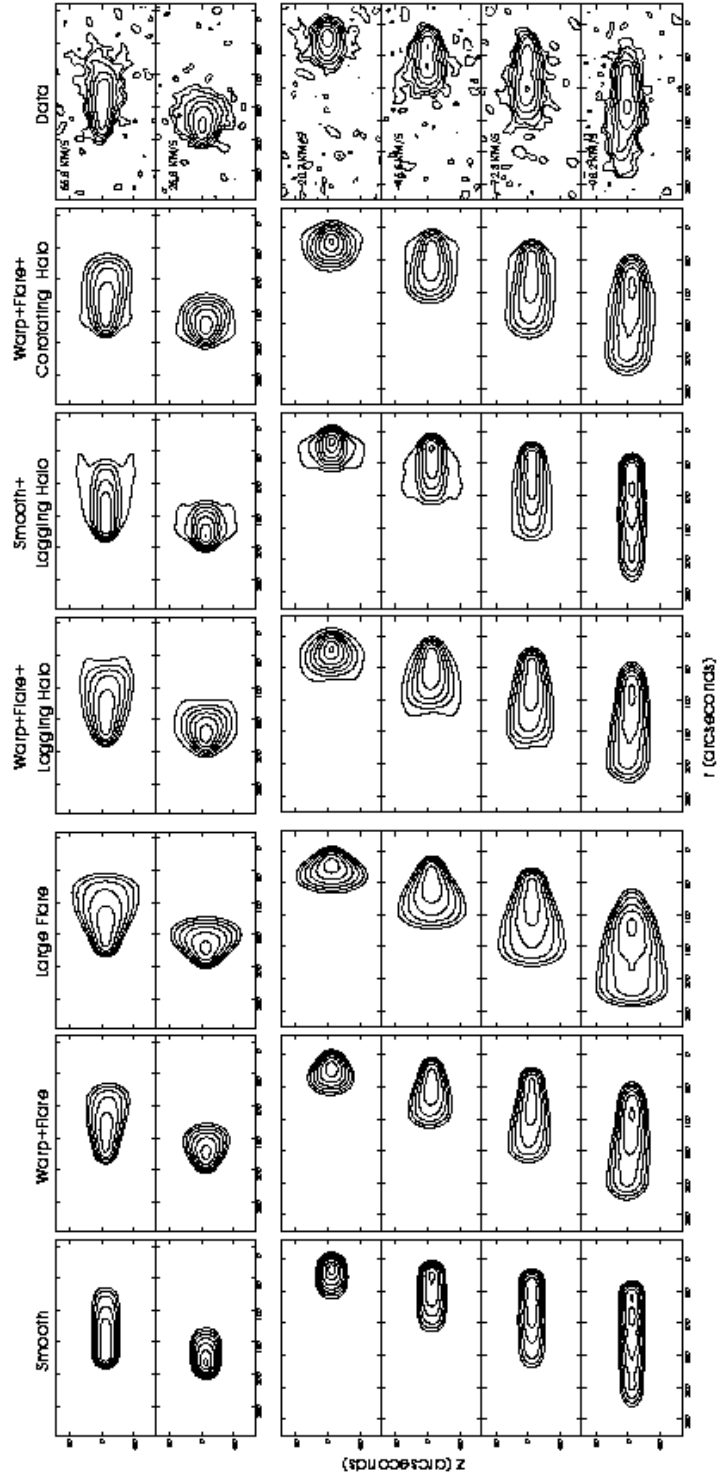}{8.0in}{0}{90}{90}{-300}{-60}
\figcaption{Representative \HI\ channel maps from our UGC~7321 VLA data
(righthand column) and several of the galaxy models discussed in
Section~\ref{3Dmod}. 
The upper two panels in each column are from the
receding side of the disk, while the lower four panels are from the
approaching side. Contour 
levels are 
-0.4,-0.2,0.2,0.4...6.4,10$\times$(0.39 mJy beam$^{-1}$)
(spatial resolution $\sim16''$ and velocity resolution $\sim$5.2~\kms), 
with the
lowest contour approximately 2$\sigma$.
\protect\label{fig:modchanmaps}}
\end{figure}

For the smooth model, we found the best overall fit to the data
using a FWHM \HI\ thickness of 370~pc,  and a central density 
of $\sim$0.19~cm$^{-3}$. 
In Figure~\ref{fig:modchanmaps} we plot several of the
channel maps produced
from this model alongside the  real data\footnote{The full set of
channel maps for UGC~7321 is presented in Uson \& Matthews 2003.}.  In
Figure~\ref{fig:modZPV} we plot several representative
minor axis P-V plots for this model.

We see from Figure~\ref{fig:modchanmaps} that the distribution
of the emission in most of the smooth model channel maps is not a
good match to the data;
in some of the channels,  the model emission 
narrows in $z$-extent with increasing $r$ (e.g., the $-72.3$ and
$-98.2$~\kms\
panels), while
in the actual data, the $z$-distribution of emission broadens
with increasing $r$. In addition, the smooth model cannot simultaneously
match the intensity in the midplane while still reproducing the 
full observed $z$-extent of the emission (see also Figure~\ref{fig:modZPV}). 
Finally, the mean FWHM thickness of the gas at $r=0$
inferred from this model is larger than that of the stars in the
near-infrared [(FHWM)$_{*,NIR}\approx$260~pc at $r=0$; Matthews (2000)], 
and exceeds typical \HI\ scale heights of normal spiral galaxies.
For example, the mean FWHM of the Milky Way's cool \HI\ disk at
intermediate galactocentric radii (4-8~kpc)
is only $\sim$230~pc 
(Dickey \& Lockman 1990). Since in general, all but the youngest 
stars are dynamically hotter
than the cool \HI\ disks of normal
galaxies, this smooth model is unappealing on physical
grounds, particularly for such a dynamically cold system as UGC~7321. 
Some of these problems can be partially
alleviated by modelling the emission in terms of {\sl two} disk
components of differing scale heights. However, such a model still
cannot reproduce the broadening of the gas in the $z$ direction with
increasing projected 
radius that is seen in a number of the individual channel maps.

Using our smooth model,
we have explored the additional possibility that this
model produces a poor match to the data simply because we have
overestimated the inclination of UGC~7321. However, computing a similar
model to that described above, except with $i=86^{\circ}$,  yields 
characteristic 
``V-shaped" channel maps, unlike those seen in our data (see also
Figure~4 of Swaters et al. 1997). Moreover, this $i=86^{\circ}$
model  predicts too little density
concentration along the midplane. We conclude that our adopted
inclination for UGC~7321 
is not significantly in error. 

It is clear from the \HI\ total intensity map in
Figure~\ref{fig:mom0} that the \HI\ disk of UGC~7321 has a mild
integral sign warp (see also Uson \& Matthews 2003). Therefore a
more accurate model of UGC~7321 naturally
should take this into account. In addition, inspection of the individual
channel maps of UGC~7321 suggests
that the \HI\ layer appears to be
flared (i.e., the $z$-extent of the gas increases with increasing
projected radius).
In our next set of computations,
we therefore introduced disk {\sl warping} and {\sl flaring} as follows:

\begin{equation}
\rho_{HI} \propto \exp[{-0.5(r/h_{r})^2}]
 \exp[{-0.5(z'/h_f)^2]}\; .
\end{equation}
The flaring scaleheight is
$h_f = h_1+h_{f0}r^{5/4}$.  The warp is introduced by setting
$z'= |z-z_{\rm warp}|$, $z_{\rm warp}=z_{w}\cos{\phi}$ with $\phi$
the azimuthal angle in the disk midplane, and
$z_w = z_{w0}(r-r_w)^{5/4}$ for $r > r_w$.

This analytic form for the flare was
adopted based on its ability to provide a good fit to the
data. Models  where the scale height remained nearly
constant within the inner few kpc (as is inferred, for example, in the
Milky Way; Burton 1988) produced noticeably poorer fits. 

The
radius where the warp commences was taken to be 5.8~kpc (the edge of
the stellar disk), 
and its maximum amplitude to be 480~pc
(i.e., $r_w=5.8$~kpc and $z_{w0}=0.05$~kpc). We assume
that the line of nodes of the warp is straight throughout the
disk.  For the flaring, we found our best overall fit to the data taking
the FWHM of the disk
to be 260~pc
at its center and taking the maximum thickness at $r_{max}$ to be 
1.64~kpc FWHM.
The corresponding scaleheight parameters 
are $h_1=0.11$~kpc and $h_{f0}=0.027$~kpc. 

Although it is difficult to accurately constrain gas scale heights at
small galactocentric radius for edge-on disks (e.g., Olling 1997a,b),
we found that models with much smaller gas scale 
heights at $r$=0 provided less acceptable  
fits to the data. This would imply that the
scale height of the cool \HI\ disk in UGC~7321 is  not significantly smaller
than that of the stars.
In fact, we note that with our adopted
parameters, in order to properly fit the outer \HI\ disk scale height,
we find that the resulting FWHM of the gas layer in our best model 
actually exceeds by $\sim20$\% that of the
stars as measured in the NIR  near the outskirts of the
stellar disk (Matthews 2000). The requirement that $h_{gas}< h_{*}$
therefore provides some
estimate of the uncertainty in our flare model
parameters. Interestingly, a near-equality of the gaseous and stellar
scale heights in the outer disk of UGC~7321 was also suggested by
Gallagher \& Matthews (2002) based on the vertical distribution of luminous
outer disk stars and dark nebulae seen in {\it Hubble Space Telescope} images.

Our resulting
warp+flare  models are illustrated in 
Figure~\ref{fig:modZPV} and Figure~\ref{fig:modchanmaps}.
Figures~\ref{fig:modZPV} \& \ref{fig:modchanmaps} show that our warp+flare
model results is a substantial improvement in fitting the 
channel maps and the minor axis P-V plots. Overall this model reproduces 
quite well the
bulk of the brightest midplane emission in UGC~7321. However, we find that
the warp+flare model still is unable 
to reproduce the more diffuse, highest $z$-height emission. This
shortcoming is most evident for channels corresponding to 
velocities within $\sim$60~\kms\ of systemic.
Furthermore, one-dimensional
intensity cuts extracted in a manner 
analogous to those in Figure~\ref{fig:1Dcuts}
do not exhibit the same type of extended wings seen in the data;
rather for the warp+flare model, the intensity distribution can be almost
completely reproduced by a single (albeit broader) Gaussian
(Figure~\ref{fig:mod1D}). 

Constraining the peak 
amplitude 
of the flare to match the maximum observed $z$-extents of the gas
(i.e., FWHM of the flare equal to $\sim$3.3~kpc at $r_{max}$) significantly
lessens the agreement with the data, as the highest $z$-height
flared material does not
follow either the spatial distribution or the velocity structure
of the observed high latitude
gas (viz. the ``large flare model'' in
Figures~\ref{fig:modZPV} \& \ref{fig:modchanmaps}; note
particularly the $-72.3$ and $-98.2$~\kms\ channel maps). 
Moreover, models with a higher amplitude flare fail to
match the observed total \HI\ intensity distribution since the disk
begins to appear thickest at its edges, contrary to what
is observed (Figure~\ref{fig:modmom0}).

\subsubsection{Two-Component Disk Models\protect\label{twocomp}}
In our own Galaxy, current evidence suggests that within the solar radius,
\HI\ material with $|z|\le$1 kpc (or possibly as high as $|z|\le$3~kpc)
corotates with the gas in the
plane (Dickey \& Lockman 1990; Benjamin 2000; Lockman 2002). 
However, alternate tracers of the gaseous halo at higher $z$-heights (e.g.,
\CIV) have shown
some evidence for lagging rotation within the inner Galaxy
(Savage \& Massa 1987). In external galaxies, evidence of lagging
halos of ionized gas has also been reported 
(Rand 1997,2000; T\"ullmann et al. 2000). 
In addition, as noted above, 
evidence for rotationally lagging \HI\ halos
has now been found for a few other carefully studied spirals. 
We therefore explore the possibility that
such a model might also describe the high-latitude \HI\ emission structure
in UGC~7321.

For these ``lagging halo'' models we have included the following two
components:
(1) a thin disk of material that rotates according to the disk
rotation curve and
whose parameters (except for the central density) 
are fixed to those derived for our best-fitting warp+flare model
above; (2) a thicker, Gaussian-shaped layer whose rotational 
velocity is permitted
to decrease linearly as a function
of $z$-height. 
We use a velocity dispersion for the ``halo'' component of 
$\sigma_{HI,halo}=\sigma_{HI,thin}(h_{z,halo}/\bar h_{z,thin})$
(e.g., Kulkarni
\& Fich 1985; Combes \&
Becquaert 1997). Here $\sigma_{HI}$ is the velocity dispersion in the
thin disk (7~\kms\ in all models) and $h_{z,halo}$ 
and $\bar h_{z,thin}$
are the scale height of the halo and the mean scale height of the
flared thin disk, respectively.

\begin{figure}
\plotfiddle{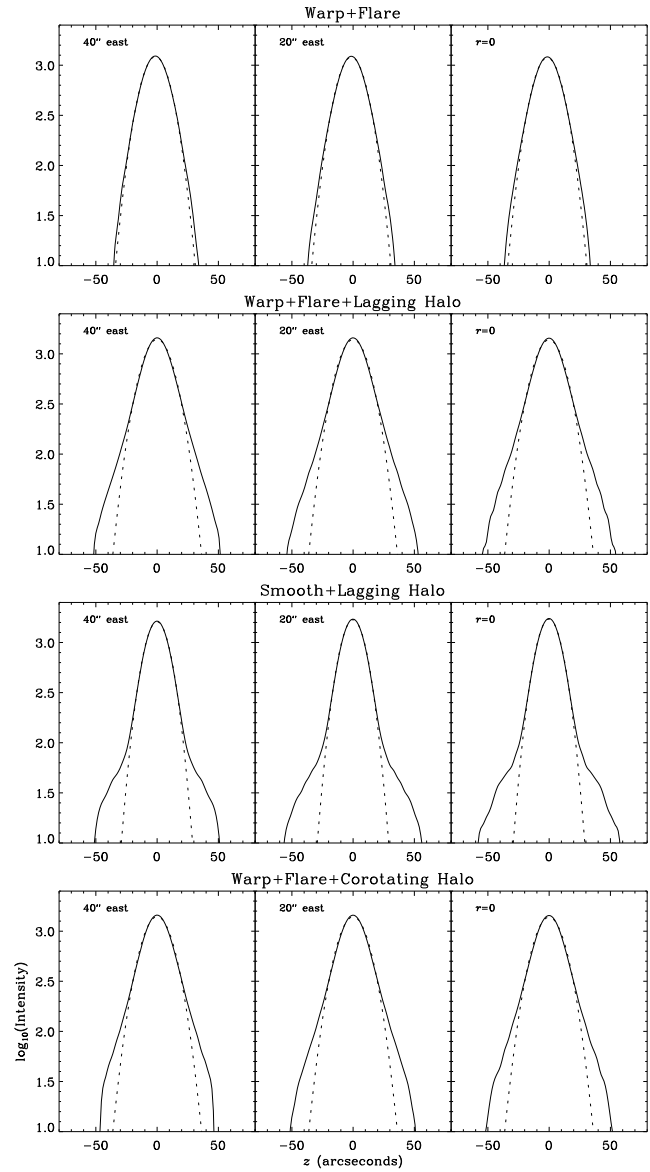}{8.0in}{90}{90}{90}{510}{0}
\figcaption{Same as Figure~\ref{fig:1Dcuts}, but for the
``warp+flare'', 
``warp+flare+lagging halo'', ``smooth+lagging halo'', and
``warp+flare+corotating halo''
models.\protect\label{fig:mod1D}}
\end{figure}

\begin{figure}
\plotfiddle{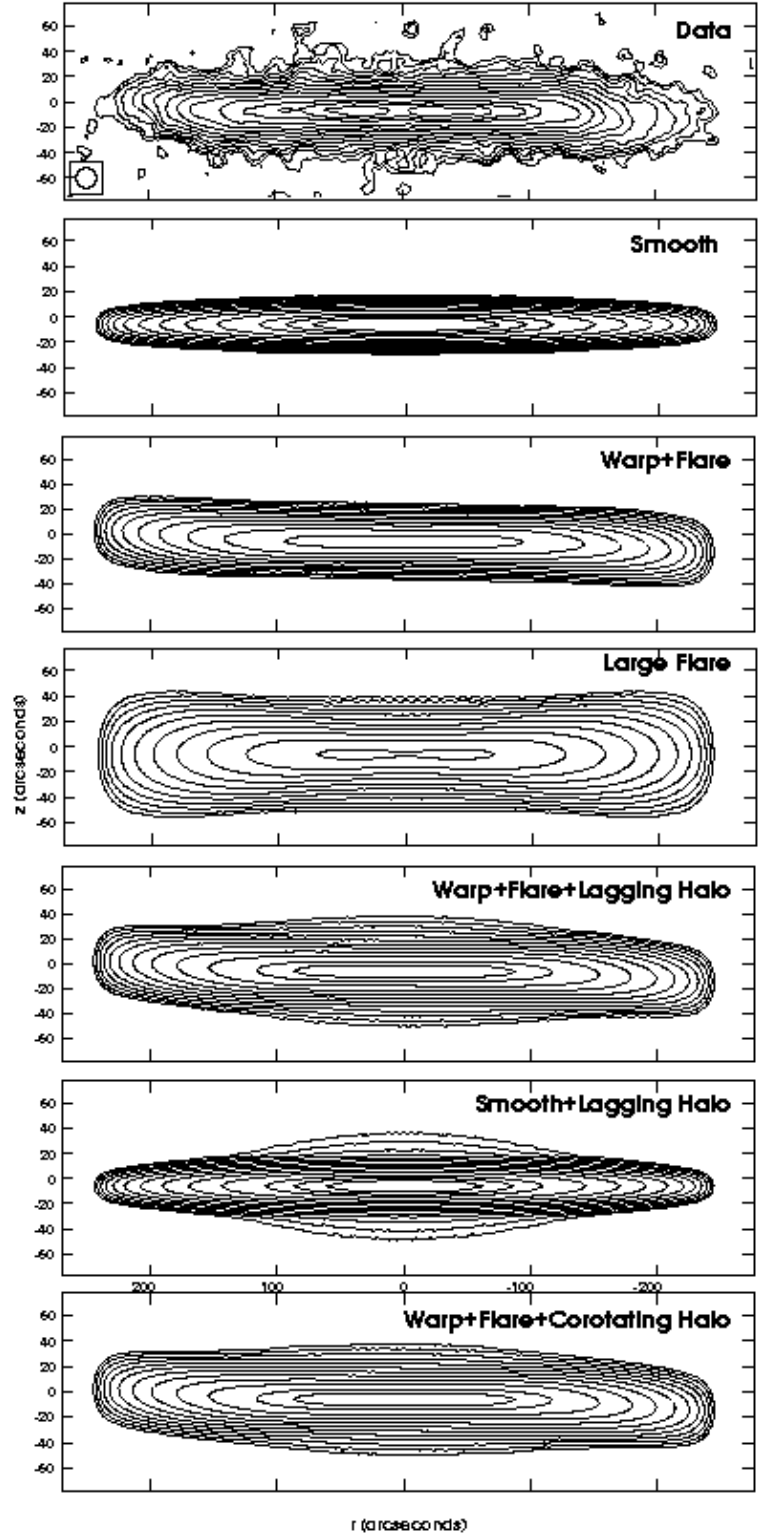}{8.0in}{0}{100}{100}{-300}{-110}
\figcaption{\HI\ total intensity maps for the various galaxy models
discussed in Section~\ref{3Dmod} (lower panels) and for our data
(upper panel).  Contour levels are
1.4,2,2.8,4,...44,64,88$\times(7.7\times10^{19}$) atoms
cm$^{-2}$ (spatial resolution$\sim16''$ and velocity 
resolution $\sim$5.2~\kms). Only points in each channel map 
with flux $>-1.5\sigma$
were included when computing in these images. 
\protect\label{fig:modmom0}}
\end{figure}

In Figures~\ref{fig:modZPV} and \ref{fig:modchanmaps} 
we show the results for a model where the halo component has the
following parameters: a FWHM
of 3.3~kpc, a central volume density 16 times lower than that of the
disk material, a velocity dispersion of 35~\kms, 
and a velocity that declines linearly to 50\% of the in-disk
value at $|z|=$2~kpc, and remains constant thereafter. 
This corresponds to a falloff of
$dV(r)/dz\le$25~\kms~kpc$^{-1}$. While the
particular combination of model parameters we have used
is not necessarily unique, we see from
these figures that such a model reasonably reproduces the
spatial and velocity
distribution of \HI\ as observed in UGC~7321. The total intensity
distribution derived from this model also matches well with the data
(Figures~\ref{fig:mod1D} \& \ref{fig:modmom0}). 

Given the success of the lagging halo model at reproducing the \HI\
structure of UGC~7321, we also explored whether the emission could be
reproduced with a smooth disk and a lagging halo alone, without
any flaring (the ``smooth+lagging halo'' model). 
The channel maps and vertical P-V slices predicted
by such a model provide fair agreement with the data
(Figures~ \ref{fig:modZPV}  \& \ref{fig:modchanmaps}), although noticeable
discrepancies begin to occur in the $-98.2$~\kms\ channel map in
Figure~\ref{fig:modchanmaps} and the \am{3}{4}W P-V slice in
Figure~\ref{fig:modZPV}. Moreover, such a model clearly is seen to break
down in Figures~\ref{fig:mod1D} \& \ref{fig:modmom0}.
One-dimensional vertical intensity slices from this model show 
more pronounced non-Gaussian wings than does the warp+flare+lagging
halo model. And the
total intensity image created from this model is seen to significantly
underestimate the thickness of the disk at large projected radii.
We conclude that UGC~7321 cannot be modelled without including a
flaring of the gas layer.

Lastly, we have also tested a model that includes warping and flaring, but
where the halo {\sl corotates} with the disk (the
``warp+flare+corotating halo'' model). As seen in
Figures~\ref{fig:modZPV} \& \ref{fig:modchanmaps}, in several 
channels this  model seems to
predict too much high-latitude emission at small projected radii (compare 
the $V=$56.8, $-$46.5, 
and $-98.2$~\kms\ channel maps in
Figure~\ref{fig:modchanmaps}). However, only the outermost
($\sim2\sigma$) contours are affected, and it is only at these levels
that the differences between the lagging and corotating models
become readily apparent. 
To complicate matters further, 
in some instances, vertically extended
gas is seen at locations more consistent with the corotating model than
the lagging halo model (for example, the vertical P-V cut at
\am{1}{9}~W in Figure~\ref{fig:modZPV}).
This could reflect noncircular motions of the emitting gas, or may
indicate that the rotational lag of the halo has a dependence on $r$
different from that in our models. At
present, we are unable to distinguish between these possibilities.

\subsubsection{Tests of the Models in the Presence of
Noise\protect\label{noisetests}}
\begin{figure}
\plotfiddle{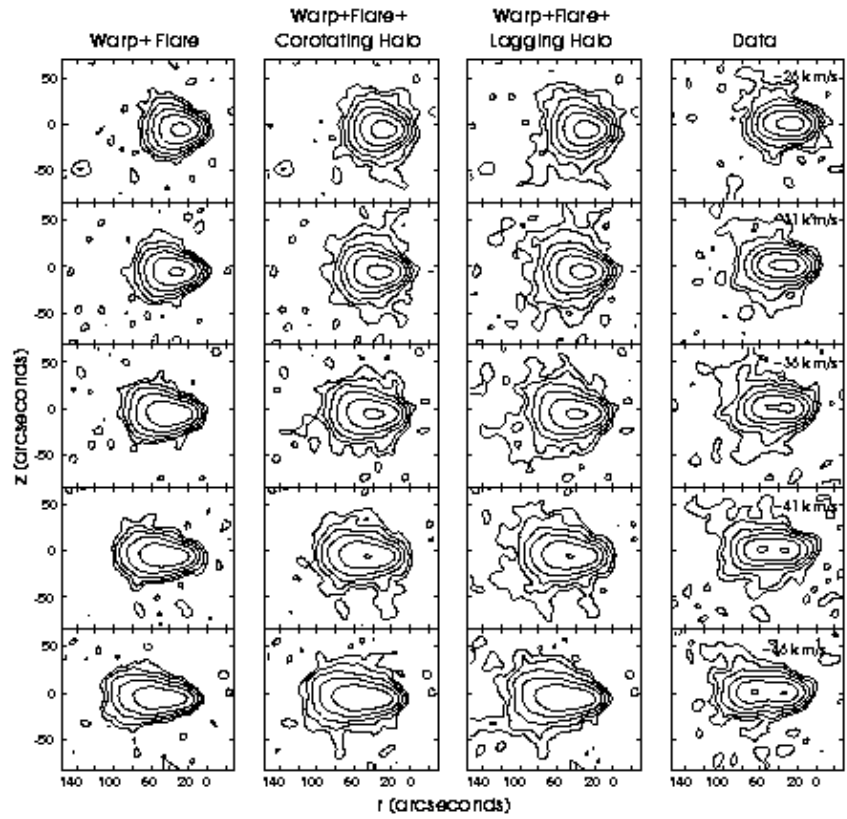}{8.0in}{-90}{100}{100}{-400}{600}
\figcaption{Selected channel maps from three of our models with
Gaussian noise added (see Section~\ref{noisetests}). 
The corresponding channel maps from
the real data are also
shown.
Contour levels are as in Figure~\ref{fig:modchanmaps}.
\protect\label{fig:noisycmaps}}
\end{figure}

\begin{figure}
\plotfiddle{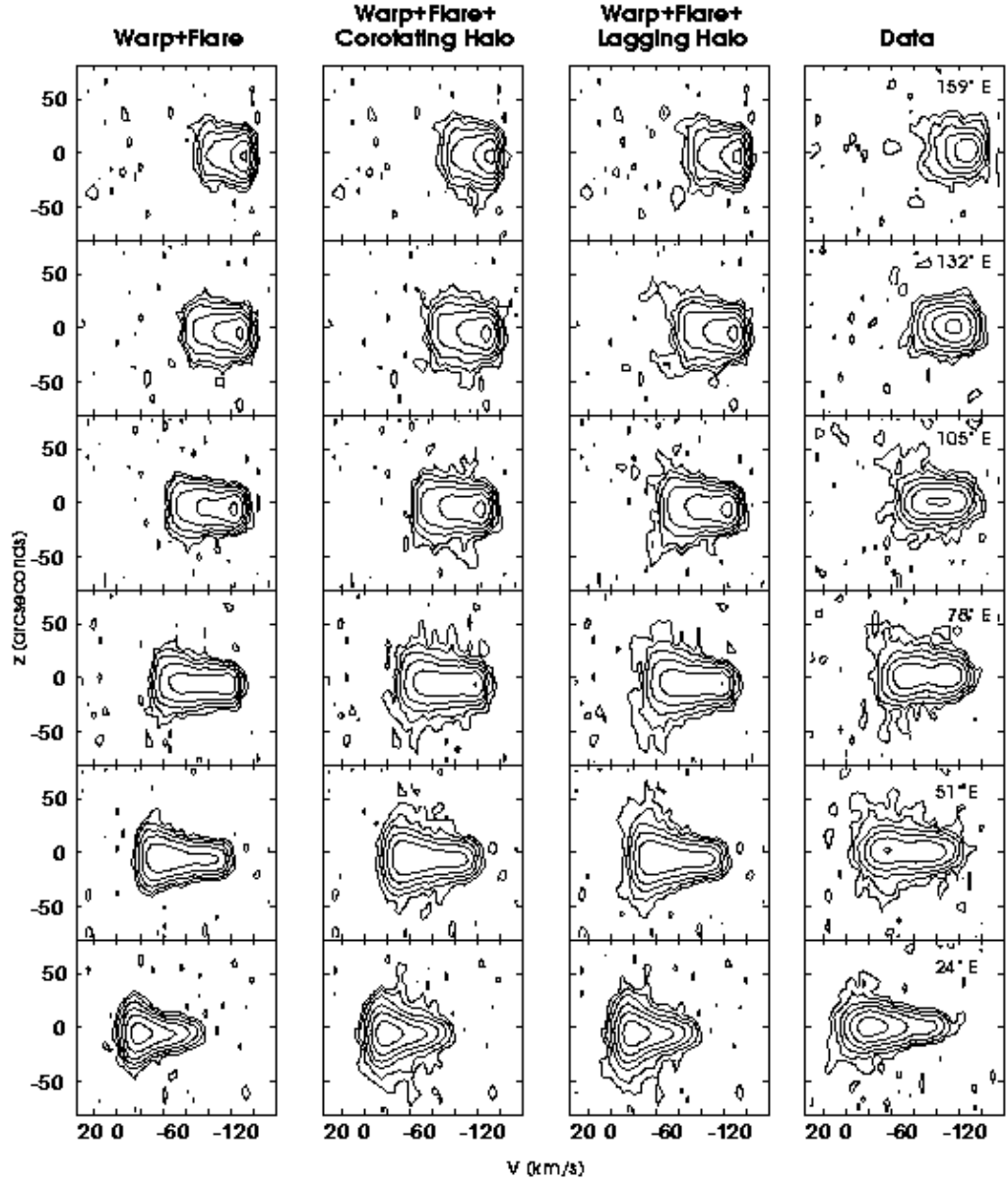}{8.0in}{-90}{100}{100}{-400}{600}
\figcaption{Position-velocity profiles extracted parallel to the minor
axis at several disk locations for three of our models, with Gaussian
noise added. The real data at the corresponding positions 
are also shown.  The profiles in each row were
extracted at 27$''$ intervals between 159$''$ east and 24$''$ east of
the minor axis. Contour
levels 
are as in Figure~\ref{fig:modZPV}.\protect\label{fig:noisyZPV}}
\end{figure}

To better gauge how well we can distinguish between a lagging versus a
corotating halo model given the noise in our observations, we have added to
our models Gaussian noise identical to that in the real data and
then re-examined the resulting channel maps and vertical P-V
slices. We repeated this exercise several times in order to gauge how
much faint features will appear to ``vary'' in the presence of random noise. 
We have also performed a similar experiment on the pure warp+flare
model. Examples of channel maps and vertical P-V slices from
these ``noisy'' models are
shown in Figures~\ref{fig:noisycmaps} \& \ref{fig:noisyZPV}.

From Figures~\ref{fig:noisycmaps} \& \ref{fig:noisyZPV} we can clearly
see that noise alone is insufficient to emulate the types of
high-latitude features we see in UGC~7321, as the warp+flare model
shows essentially no evidence for such features. We also find that some
differences between the lagging and corotating models do
persist even in the presence of noise.
For example, in the channel maps over the velocity range
20-50~\kms,
the lagging model consistently shows high-latitude material
``trailing'' to large projected radius, 
behind the bulk of the low-latitude emission. Such features are consistently
seen in the data, but are not reproduced by the corotating halo model.
Overall, at intermediate radii, the lagging halo model predicts more
high-$z$ emission near the systemic velocity than does the corotating
halo model, and this also seems borne out by the data at most disk
locations where high-$z$ \HI\ is observed (e.g., the 51$''$, 78$''$, and
105$''$E P-V slices in Figure~\ref{fig:noisyZPV}).

Based on the above arguments, we suggest a tentative preference for the
lagging halo model for UGC~7321 over one where the halo corotates with
the disk, with the caveat that a lagging halo model would seem to
require the addition of 
some non-circular motions and/or a possible additional
change in $V(z)$ as a function of $r$. However,
given the limitations of present data and models, a situation
where the bulk of the high-latitude
gas corotates with the disk cannot yet be strongly ruled out.

A further outcome of the exercise of introducing noise into our models
is to illustrate that a smooth \HI\
halo can appear to break up into semi-discrete features in the
presence of noise. Thus while the high-latitude gas in UGC~7321 may in fact
have an intrinsically 
clumped or filamentary nature, this becomes difficult to distinguish
from a relatively smooth emission distribution in our present data.

\section{Discussion: the Origin of the Vertically Extended HI 
Emission\protect\label{discussion}}
We have shown in the previous sections 
that the LSB spiral UGC~7321 contains  vertically
extended \HI\ gas whose $z$-height extends beyond that predicted
for a simple, cool \HI\ layer, even after accounting for the effects
of warping and 
flaring. Moreover, based on our best-fitting 3-D models, we have
suggested that at least a fraction of this high-latitude material
may lag in rotation relative to the material in
the midplane. The caveats are that such a model would seem to
require the addition of 
some non-circular motions and/or a possible additional
change in $V(z)$ as a function of $r$; also, 
given the limitations of our data and models, a situation
where the bulk of the high-latitude
gas corotates with the disk cannot yet be definitively ruled out. 
The
emergence of a corotating halo 
model would have interesting implications for the
origin of the high-latitude 
gas in UGC~7321, since both ballistic models (i.e., those associated
with galactic fountain flows) and hydrostatic models for galactic
halos  predict rotational lags. In the case of the hydrostatic
models, cylindrical rotation can be achieved only if there exist additional
pressure gradients or magnetic tension in the halo (Benjamin 2002).

\subsection{Inferred Properties of the HI Halo\protect\label{haloprop}}
Independent of the exact kinematic structure of the high-latitude gas
in UGC~7321, we can use our models to
compute an estimate of the total mass contained in
the high-latitude \HI\ component in UGC~7321.
For either our lagging or corotating halo model, the total mass of
\HI\ in the halo component is $\sim1.3\times10^{8}~M_{\odot}$
(roughly 12\% of the total \HI\ content of UGC~7321). Of this material,
$\sim1\times10^{7}~M_{\odot}$ is present at
$|z|\ge30''$ (roughly 1\% of the \HI\
content of UGC~7321). This latter value is approximately 
fifteen times less than the fractional \HI\
content inferred to be at comparable $z$-heights in NGC~891 by Swaters et
al. (1997). 
We now
briefly consider possibilities for the origin of this (comparatively weak)
halo component in UGC~7321.

\subsection{Supernova Heating/Galactic Fountain\protect\label{SN}}
For those nearby galaxies whose gaseous halos have been studied to date,
there appear to be well-established links between the
properties of the multi-phase
halos and the current star formation rates of the galaxies,
implying the two are linked (e.g., Dahlem 1997 and references therein). This
suggests that at least some  
portion of the halos may be formed
via internal processes.

The type of thickened, rotationally lagging 
\HI\ layer that we infer for UGC~7321 in Section~\ref{3Dmod} 
is qualitatively similar to the
expectation for material originating via
a galactic fountain, as described by Bregman (1980). 
In this scenario, gas
heated by supernovae is predicted to move both toward larger $|z|$ and
larger $r$ before cooling, recondensing, and returning near the point of
origin via ballistic orbits. While in the halo, the
gas should lag the disk rotation as a consequence of conservation of angular
momentum. Using qualitative arguments, 
Swaters et al. (1997) and Schaap et al. (2000) have suggested
scenarios of this type may 
account for the origin of the \HI\ halos of NGC~891
and NGC~2403, respectively. 
However, it is unclear whether a detectable \HI\ halo could be
formed in this manner in a galaxy like UGC~7321, that exhibits
such modest levels of current star formation. 

Based 
on the {\it IRAS} far-infrared 60 and 100$\mu$m fluxes for UGC~7321
(0.344~Jy$\pm$14\% and 0.964~Jy$\pm$19\%,
respectively)\footnote{Values taken from the NED database.} and the
formulae given by Condon (1992), we estimate the current
star formation rate in UGC~7321 for stars with $M\ge5$~\msun\ is only
$\sim$0.006~\msun\ per year. This is $\sim$200 times smaller than the value
similarly derived for NGC~891. 
On the other hand,
UGC~7321 is a smaller, lower mass spiral than
others where high-latitude \HI\ has been reported, and likely also has
a more tenuous interstellar medium. These factors may increase the
efficiency with which supernovae can deposit gas at high latitudes. 

An independent estimate of the star formation rate in UGC~7321 can be
derived from the H$\alpha$ data of Matthews et al. (1999).
Although the H$\alpha$+[\NII] image presented by these authors is not flux
calibrated, using the known $R$-band luminosity of the galaxy, 
the mean H$\alpha$/[\NII] line ratio from Goad \& Roberts (1981), and
the relative throughputs of the broad-band continuum ($R$)
and narrow-band (H$\alpha$+[\NII]) filters used by Matthews et al.,
we estimate an H$\alpha$ luminosity for UGC~7321 from their data of
$L_{H\alpha}\sim1\times10^{40}$~erg~s$^{-1}$. Such a value is
typical of quiescent, late-type spirals (e.g., Thilker et al. 2002).
Using the formula given by Condon (1992), this corresponds to a
massive star formation rate of $\sim$0.02~\msun\ per year--a few times
higher than estimated above from the FIR emission, but roughly
consistent to within the uncertainties of both estimates.

In order to test whether it is feasible that supernovae could supply
the energy required to maintain an observable \HI\ halo in UGC~7321,
we follow an approach similar to Kulkarni \& Fich (1985). 
We neglect possible partial
support for the gas that may come 
from magnetic fields, cosmic rays, and radiation, since 
based on the extremely weak radio
continuum flux from this galaxy (Uson \& Matthews 2003), these terms 
are likely to be quite small.
In this case, assuming hydrostatic equilibrium, 
the kinetic energy of the halo is simply given by
$E_{k,halo}=\frac{1}{2}M_{halo}\sigma^2_{HI,halo}$, 
where $M_{halo}$
is the mass of the gaseous halo material (\HI\ plus He) 
and $\sigma_{HI,halo}$
is the velocity dispersion of the halo (35~\kms\ for our model). 
Using the halo mass
quoted in Section~\ref{haloprop}, applying 
a factor of 1.34 to correct for He, and
assuming a radius for the galaxy of 11.8~kpc, we derive 
the kinetic energy per
unit area of 
$E_{k,halo}/A=5.1\times10^{8}$ erg cm$^{-2}$ for the gaseous halo.

We assume that the timescale for intercloud collisions in the halo
is likely to be
longer than their free-fall time back to the disk, and therefore use
the latter timescale 
to estimate the rate at which energy must be continuously
supplied to maintain the halo. Assuming a downward velocity of 35~\kms\
from a height of 1.6~kpc, the approximate free-fall time is 
$t_{ff}\sim$46 million years. This then  requires an energy input rate of
$E_{k,halo}/(At_{ff})\approx3.6\times10^{-7}$ erg s$^{-1}$
cm$^{-2}$. 

From Cappellaro et al (1993), we find that
mean supernova rates for typical Scd/Sd galaxies are 
2.43$\pm$0.70 per 100 years, per
$10^{10}L_{\odot}$, of which $\sim$24\% are typically type Ia 
and $\sim$76\% are
core collapse supernovae. Since the $B$-band luminosity of UGC~7321 is
1.1$\times10^{9}L_{\odot}$ (Uson \& Matthews 2003), this yields an estimated
total supernova rate of $\sim3\times10^{-3}$ year$^{-1}$. This number
is roughly a factor of two 
higher than we might estimate from the massive star formation
rate derived from the FIR fluxes above, but given the large uncertainty in
the FIR measurements, we adopt this as a more
``optimistic'' estimate.

Fiducial kinetic energies for type Ia and core collapse supernovae 
are $5\times10^{50}$ ergs and $10^{51}$ ergs,
respectively (Abbott 1982). Of this energy, approximately 
1\% and 3\%, respectively can be converted into bulk motions of gas
clouds (Spitzer 1978). This predicts a kinetic energy per unit
area supplied by supernovae 
of ${\dot E}_{SN}/A\sim5.0\times10^{-7}$ erg s$^{-1}$ cm$^{-2}$---a
value comparable to ${\dot E_{k,halo}}/(At_{ff})$ estimated above, and 
therefore consistent with the possibility
that supernova energies may be sufficient to maintain the
high-latitude \HI\ observed in UGC~7321. 

We can provide an independent check on our energy injection
estimate  using
the figures and relations supplied by Collins et al. (2002). 
Using their Figure~4, we find that a maximum observed gas height
of 2.4~kpc implies a ratio of ``kick'' velocity $V_{k}$ to circular velocity
$V_{c}$ 
in the midplane of $\sim$0.8. Choosing some intermediate galactocentric
radius (say 4~kpc), and taking $V(r)\approx$100~\kms\ for UGC~7321,
this implies
$V_{k}$=80~\kms. From Figure~6 of Collins et al., the implied ``cycling
frequency'' is $f_{cycle}=1.4\times10^{-8}$yr$^{-1}$. Inserting this
into their Equation~5 leads to an estimate of the halo mass flux of
${\dot M}_{h}\sim2.3~M_{\odot}$~yr$^{-1}$. Finally, from the relation
for the energy input requirement: ${\dot
E}=(1.1\times10^{39}{\rm erg s^{-1}}){\dot
M}_{h}V^{2}_{k,100}$ (where the halo mass flux is in
units of $M_{\odot}$~yr$^{-1}$ and the kick velocity is in units of
100~\kms) we arrive at ${\dot E}=1.6\times10^{39}$~erg~s$^{-1}$, or
${\dot E}/A=3.8\times10^{-7}$~erg~s$^{-1}$~cm$^{-2}$---very close
to our
estimate above.

Using the ``kick'' velocity estimated from the Collins et al. models,
we can make another interesting comparison, namely how that value
compares with the escape speed from the galaxy, $v_{e}$. Assuming a
spherical potential, the escape speed from a galaxy is given by equation
2-292 of Binney \& Tremaine (1987). At our fiducial radius of
$r=$4~kpc, we thus derive an escape speed
$v_{e}\approx$141~\kms---significantly higher than the kick
velocity of 80~\kms, implying UGC~7321 would 
be able to retain most of its
fountain gas. 

Of course the above estimates are highly
idealized, and
more sophisticated computations are clearly needed to test more
accurately how efficiently
supernovae (and stellar winds) can be
in supplying gas to the halo, and whether the observed spatial and
velocity  structures
of the high-latitude material 
quantitatively matches the predictions of galactic
fountain-type models. Very deep
H$\alpha$ observations to search for an extraplanar component of 
ionized gas in
UGC~7321 would also be of interest in further testing the link between star
formation and high-latitude \HI\ in this galaxy.  
In the mean time, we also consider the feasibility of other
possibilities for the origin of the halo material in UGC~7321.

\subsection{Alternative Possibilities}
In spite of the well-established links between current star formation and the
existence of gaseous halos in brighter galaxies, it
is possible that star formation simply acts to 
augment the halo and
lead to the partial ionization of pre-existing 
halo gas, thereby making these halos more extensive and more readily
observable via a variety of tracers. This may be expected to occur
if some gas were deposited in the halo either during an earlier phase in the
galaxy's evolution, or via ongoing 
accretion. We now consider these two
possibilities in turn for the source of the high-latitude \HI\ in UGC~7321.

\subsubsection{Galactic Superwinds}
Steidel et al. (2002) recently reported that
for a sample of five intermediate redshift \MgII\ absorption line
galaxies, the models needed to explain the kinematic properties of the
absorbers require either thick rotating gas layers and/or rotational
velocities that decline with height above the plane. One galaxy in
their sample, G1~1222+228, is particularly interesting in light of our
findings for UGC~7321, as it appears
to show some intriguing analogies. 

Like UGC~7321, G1~1222+228
(redshift 0.55) 
has the morphology of a pure disk system of relatively low
surface brightness, in spite of its edge-on geometry. Its absolute blue
magnitude ($-18.09$) and its peak rotational velocity 
(100~\kms) both are also very similar  to UGC~7321. From Steidel et al.'s
Figure~4, we estimate the linear diameter of G1~1222+228 to be
$\sim$12.1$h^{-1}$~kpc---again comparable to UGC~7321, particularly if
the faint outskirts of the disk
are invisible owing  to their LSB nature. For this
system, Steidel et al. successfully reproduced 
the \MgII\ absorption profile for this galaxy using a model
with an ``effective height'' for the gas
layer (as measured from the midplane) of $\le$5~kpc and a velocity
structure for the high-$z$-height material that falls to zero velocity
within a few kpc above the disk plane. This model is interestingly similar
to our lagging halo model for UGC~7321.

Steidel et al. (2002) note the general problem of explaining \MgII\
halos of moderate redshift 
galaxies from galactic fountains, since most \MgII\  absorption
systems are not currently highly active star-formers. Motivated by
results from the strong far-UV absorption lines observed in Lyman
break galaxies, they suggest as
an alternative, large-scale galactic winds from a much earlier
epoch of star formation (at redshifts $\gsim$3). In this scenario,
the shock-heated and metal-enriched 
superwind material would cool slowly to form the gaseous disk and halo
observed in normal disk galaxies at low or moderate
redshifts. Could this be the origin of the high-latitude \HI\ in
UGC~7321?

Although the galactic superwind scenario suggested by Steidel et al. (2002)
has not yet been quantified in detail,
based on dynamical arguments it is not clear whether material deposited in
the halo at high redshift would still be
observable in local galaxies.  The metal-line
absorption features in the halos of galaxies
probed by QSO absorption
lines,  including \MgII, appear to arise in 
discrete clouds (e.g., Petitjean \& Bergeron 1990), and
Chen, Lanzetta, \& Webb (2001a) have argued that
the timescale for collisions and subsequent orbital decay of
these clouds should be considerably less than a Hubble
time. Therefore it is uncertain whether any remnant of
gaseous halos formed by superwinds at high redshifts could
persist to the present epoch. 

\subsubsection{Infalling Gas}
If the high-latitude \HI\ 
material in UGC~7321 is not directly tied to current
star formation, and is not a remnant from an early galactic superwind,
an alternative is that at least some fraction of 
this gas
has been acquired through the accretion of intergalactic clouds or
small companions (e.g., Oort 1970; van der Hulst \& Sancisi 1988). 
This could have occurred in the form of a one-time accretion of a
dwarf satellite or the ``drizzle'' of small parcels of \HI\
onto the galaxy.

UGC~7321 appears to be a very isolated galaxy, and Uson \& Matthews
(2003) put a 10$\sigma$ upper limit 
on the \HI\ masses of any gas-rich companions or clumps 
within 12 arcminutes (36 kpc) of
UGC~7321  of $<2.2\times10^{6}$~\msun. In addition, the mass of
any possible infalling
clumps would be limited by the constraint that they cannot excessively
heat and disrupt the disk (e.g., T\'oth \& Ostriker 1992). 

We can obtain a rough estimate of the maximum mass of a satellite, $M_{s}$, 
that could have been recently
accreted by UGC~7321 without destroying its thin stellar disk using the
relation given by Lacey (1991): 
$$M_{s}\lsim\sigma^{2}_{z,*}M_{D}V^{-2}_{max}.$$ 

\noindent Here, $\sigma_{z,*}$ is the stellar velocity dispersion,
$M_{D}$ is the disk mass, and $V_{max}$ is the peak disk rotational
velocity. Taking
$M_{D}$ to be equal to the total 
estimated mass of gas and stars in the disk, 
$V_{max}$=105~\kms\ (Uson \& Matthews
2003), and
$\sigma_{z,*}$=20~\kms\ (Matthews 2000), we find
$M_{s}\lsim 7.5\times10^{7}~M_{\odot}$. This is comparable to our
inferred halo mass for UGC~7321; however, any gas-rich dwarf
galaxy is likely to have a dark matter content equal to at least a few
times its visible mass (e.g., C\^ot\'e, Carignan, \& Freeman 2000). 
Thus it appears  possible, although
unlikely that the high-latitude
material in UGC~7321 originated from a single accretion event.

Still another possibility is that UGC~7321 has been accreting small
$M_{HI}<2\times10^{6}M_{\odot}$ parcels of gas throughout much of its
lifetime. Indeed, a variety of lines of evidence suggest that slow,
continual gas
accretion (of order a few $M_{\odot}$ per year) may be a common
occurrence for disk galaxies (e.g., Wakker et al. 1999; 
Binney 2000; see also Oort 1970).
In the
case of UGC~7321, the
misaligned angular momentum thus acquired (e.g., Binney \& Jiang 1999;
Debattista \& Sellwood 1999)
could also offer an explanation for the
origin of the warp in this isolated system (see also the 
discussion in Uson \&
Matthews 2003). In order to maintain the observed thick \HI\ layer, we
estimate that a mass infall rate of $\ga$3.8$M_{\odot}$ per year
would be required based on the relation ${\dot E}=\frac{1}{2}{\dot
M}V^{2}$, where we have taken ${\dot E}=1.5\times10^{39}$~erg~s$^{-1}$
and assumed $V\approx$35~\kms. 
This is comparable to gas infall rates estimated for
the Milky Way (Binney 2000). However, in order for 
infalling material to share 
the rotation of the disk, this requires that the clumps
must become ``dragged'' along with
some sort of pre-existing, rotating medium, and the existence of such
a medium is unclear. This objection could
partially be alleviated if infalling clumps acted
to trigger further instabilities in the disk and thus raised additional
rotating gas to high $z$.

\section{Summary}
Using sensitive VLA \HI\ observations in combination with 3-D
modelling, we have found evidence for 
high-latitude ($|z|\lsim$2.4~kpc) \HI\ emission 
in the edge-on, LSB spiral
galaxy UGC~7321.  We are unable to reproduce the spatial
and kinematic structures of this vertically extended
material solely by projection 
effects from warping and flaring of the
gas layer. However, we  are successfully able
to model the \HI\ distribution in UGC~7321 with a combination of a 
thin, cool  \HI\ disk with mild warping and flaring, 
together with a thicker ``halo'' component. 

We find tentative evidence
that the vertically extended gas exhibits a lag in rotation velocity
by of $\la$25~\kms~kpc$^{-1}$
relative to the cooler \HI\ material in the midplane. However, given
the possible presence of additional non-circular motions and the
sensitivity limits of our data, it is impossible to completely
rule out models where the
extended gas largely corotates with the disk.

We estimate a FWHM for the cool disk
component to be $\sim$260~pc at the disk center, flaring to a FWHM of 1.6~kpc
at the last observed point; we estimate 
the FWHM thickness of the halo to be $\sim$3.3~kpc.
Based on our best models, the \HI\ ``halo'' comprises roughly
12\% of the total \HI\ content of UGC~7321.
Although the best-fitting model we have derived 
for UGC~7321 is not necessarily unique, after exploring a
large range of parameter space,
we find our model to reproduce the data significantly better than other
physically motivated models, including those with a higher-amplitude
flare, or those with a vertically extended halo but no flaring.

Our detection of vertically extended \HI\ emission (beyond that attributable
to a warp or flare) in
UGC~7321 represents the first time that evidence of such an ISM component
has been reported in a
nearby galaxy with such a low current star formation rate. Our rough
calculations imply that energy input from supernovae may be
sufficient to maintain  the observed high-latitude
\HI\ in UGC~7321. However, slow, sustained gaseous  
infall (in the form of clumps with
\HI\ masses $\lsim10^{6}~M_{\odot}$) remains
another viable explanation, and could also explain the existence of
the warp in this isolated galaxy. Our findings suggest
that \HI\ halos may be common features of
spiral disk galaxies in the local universe,
and that their existence is not necessarily  
dependent on vigorous levels of current star formation.

\acknowledgements 
We are grateful to Jay Lockman, Bob Benjamin,
Mark Reid, and Hsiao-Wen Chen for valuable suggestions and discussions,
and to an anonymous referee whose comments helped to improve this work.
LDM acknowledges support from a Clay Fellowship from the
Harvard-Smithsonian Center for Astrophysics. KW is supported by a 
UK PPARC Advanced Fellowship.

\clearpage

\clearpage

\clearpage

\clearpage

\clearpage

\clearpage

\clearpage

\clearpage

\end{document}